\documentclass[11pt]{article}

\usepackage{a4wide}
\usepackage{graphicx}

\begin{document}

\begin{center}   
\textbf{\LARGE Field evidence for the upwind velocity shift at the crest of low dunes}

\vspace*{0.2cm}

P. \textsc{Claudin}$^\sharp$, G.F.S. \textsc{Wiggs}$^\star$, B. \textsc{Andreotti}$^\sharp$
\end{center}

{\small
\noindent
$^\sharp$ Laboratoire de Physique et M\'ecanique des Milieux H\'et\'erog\`enes (PMMH),
UMR 7636 CNRS -- ESPCI -- Univ. Paris Diderot -- Univ. P.M. Curie, 10 Rue Vauquelin, 75005 Paris, France.\\
$^\star$ School of Geography and the Environment,
Oxford University Centre for the Environment, South Parks Road, Oxford OX1 3QY, United Kingdom.
}

\begin{abstract}
Wind topographically forced by hills and sand dunes accelerates on the upwind (stoss) slopes and reduces on the downwind (lee) slopes. This secondary wind regime, however, possesses a subtle effect, reported here for the first time from field measurements of near-surface wind velocity over a low dune: the wind velocity close to the surface reaches its maximum \emph{upwind} of the crest. Our field-measured data show that this upwind phase shift of velocity with respect to topography is found to be in quantitative agreement with the prediction of hydrodynamical linear analysis for turbulent flows with first order closures. This effect, together with sand transport spatial relaxation, is at the origin of the mechanisms of dune initiation, instability and growth.
\end{abstract}

\section{Introduction}
\label{intro}

Field studies of desert dune dynamics commonly focus on investigating feedback mechanisms between topography, windflow forcing, spatial and temporal dynamism in sand flux, and resulting erosion/deposition at the surface \cite{LNMcKNW96,WLW96,McKNLN97,BWL11,WW12}. Such studies have endeavoured to model the contemporary mechanics of simple dune shapes (with an emphasis on barchanoid forms) from fluid mechanics principles. However, progress in our understanding has been hampered by measurement complications resulting from the distortion of flow due to acceleration and the practical difficulty of measuring shear velocity $u_*$ within the near-surface inner-layer \cite{JH75} where changes in shear stress are assumed to be significant for sand flux \cite{LWW07}. Further, there is the conceptual difficulty in applying linear fluid mechanics analysis (\cite{JH75}, \cite{S80}, \cite{HLR88}, see also review by \cite{BH98}) to dune topographies that are relatively steep and incorporate a brink. Such topography forces flow separation and generates turbulent fields that lie beyond the strict applicability of the linear theory \cite{RT81,BHA84,FRBA90,HS99,ZH79}.

In contrast, modellers have approached the question of dune dynamics from the perspective of theoretical physics, with identification and description of the different physical ingredients necessary to
reproduce dunes of different shapes \cite{KSH02,H04,KFO05,DPH10}. Research on the linear stability analysis of a flat sand bed submitted to a constant wind \cite{ACD02,ECA05,CA06,NZRC09} has revealed the importance of two key mechanisms that must occur in the case of flow over topography, and which are at the origin of dune initiation and growth. These are, (i) the upwind phase shift of the point of maximum basal shear stress with respect to dune topography and, (ii) the spatial lag in sand transport response to changing shear stress -- termed `saturation length' $L_{\rm sat}$ \cite{SKH01,ACD02}. These theoretical approaches are similar to those employed in subaqueous research on ripples and dunes \cite{K63,E70,F74,EF82,R80,McL90,C04,FCA10}.

The advantage garnered by the analytical modelling approach is that research effort is directed towards the initiation of dunes from flat surfaces, rather than toward understanding the mechanics of existing dune morphologies, as is the case with the geomorphological field studies. In this way, the modelling approach stays within the applicable limits of the linear theory, at least in the initial stages of dune development. The two key mechanisms of the upwind phase shift in velocity and the spatial sand flux lag have been largely overlooked in geomorphological field studies which have focussed on mature dune forms, which do not conform to the linear analysis. Further, geomorphological field studies have tended to focus on the measurement of broad windflow patterns at the dune-scale (establishing flow acceleration patterns, for example) rather than the subtle intricacies of the sub-metre scales of the hydrodynamic phase shift and sand flux lag.

A single wind tunnel investigation provides the only measured data for the sand flux lag \cite{ACP10}. Here the value for the lag was shown to be of the order of $1$~m. However, there are no measured data to support the hypothesis of an upwind phase shift in basal shear stress in the aeolian case. This lack of data is a hindrance to furthering our understanding of sand dune initiation and growth. Some measurements are available in the subaqueous environment from flumes (e.g.  \cite{ZCH77}, \cite{FH88}) although these are in the transitional regime between laminar and turbulent flow. Numerical calculations of the phase shift are also in this transitional regime \cite{dALB97,CNHMcL98,SDB01,YEHCKPP09}. Only one in-direct measurement of basal shear stress (established from velocity data) is available in the fully turbulent regime but this is also  derived from the subaqueous environment \cite{PKAR07}. Data investigating the structure of the flow over topography are available for the aeolian environment but these are all restricted to the non-linear regime \cite{TMB87,GI89,GTD96,FRBA90}.

In this paper we present the first field evidence of the existence of the phase shift of the point of maximum wind speed upwind of a dune crest, a theoretical requirement for dune initiation and growth. The experiment is carried out on a low dune with a 'dome' topography in order to remain within the linear regime but in fully turbulent conditions. Together with data on the spatial relaxation of sand transport in response to wind velocity variations ($L_{\rm sat}$) our measurements of the upwind phase shift are the first to substantiate these key controls at the origin of the dune instability and initiation mechanism.

\begin{figure}[t]
\centerline{\includegraphics{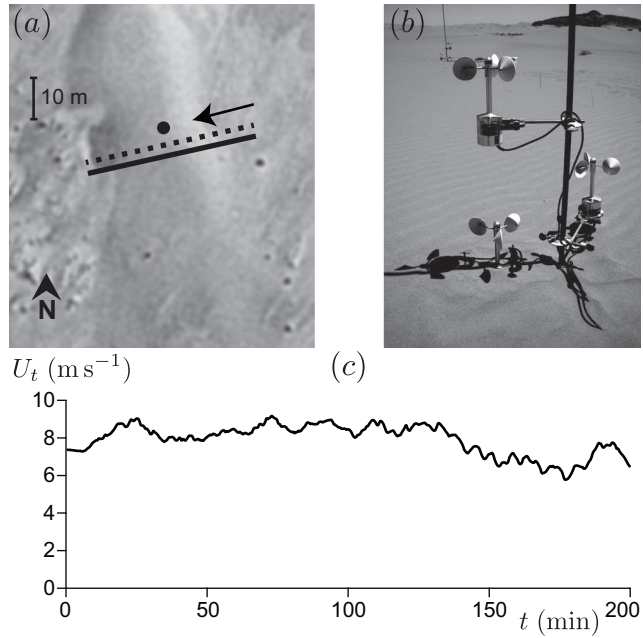}}
\caption{(a) Schematic plan view of the experimental set-up. The arrow shows the direction of the wind. Dotted line: erosion measurement transect. Solid line: wind velocity measurement transect. Bullet: location of the reference anemometer array. (b) Photo of the anemometer mast. The anemometers were located at heights of $0.11$~m, $0.30$~m and $0.50$~m with respect to the sand bed. (c) Velocity data (measured at the reference site at $0.5$~m height) throughout the duration of the experiment, averaged over a  $10$~minute moving window.}
\label{ExpSetup}
\end{figure}
%

\section{Field site and data collection}
\label{field}

Our field site was located on the western edge of the erg Oriental near Merzouga, eastern Morocco (31$^\circ$03.103~N, 04$^\circ$00.133~W), and to the south of a large star dune complex \cite{BARDHK03}. The study dune was an elongated dome, with no avalanche slip face showing that no flow separation occurred. The surrounding landscape consisted of similar dunes, all lying on flat ground, which was dry and mostly devoid of vegetation. The dune was typically $30$~m long parallel to the wind direction with a height of around $2$~m. In Fig.~\ref{ExpSetup}a, we display a schematic plan view of the dune. The dune sand had an average particle size of $310$~$\mu$m as determined by laser granulometry. The surrounding surface consisted of poorly sorted gravel in a sand matrix (grain size is mostly distributed in the range 200--500~$\mu$m). This minor change in surface roughness between the upwind plain and dune surface is not considered significant with respect to the near-surface (below $0.5$~m high) windspeed measurements described below.

We established measurement sites along the centre-line of the dune parallel to a wind blowing from a compass bearing of 74$^\circ$ between 1000 and 1400 on 23 April 2011. In order to capture the subtle variations in velocity that we were expecting in the crest area of the dune, measurement sites here were established at $0.5$~m intervals whereas sites on both the extreme upwind and downwind flanks of the dune were established at between $1$ and $8$~m intervals. The topography of the centre-line of the dune was surveyed using standard levelling techniques to an accuracy of approximately $1$~cm.

Two parallel measurement transects, approximately $1.5$~m apart, were set up (see Fig.~\ref{ExpSetup}a). One transect was dedicated to measurement of surface change with small 'brochette' sticks inserted vertically at each site into the sand to act as erosion pins. The height of each exposed pin was measured both at the beginning and end of the experiment with a tape measure to an accuracy of $\pm 1~$mm. These measurements provided data on the amount of erosion or deposition evident on the dune surface during the $4$~hours of the experiment and, when integrated over the distance between pins, provided an assessment of sand flux.

The second measurement transect was dedicated to wind velocity measurements. Two anemometer arrays were deployed, each consisting of three switching cup anemometers (Vector A-100R; accuracy $0.1$~m s$^{-1}$, distance constant $2.3$~m) at heights of $0.11$~m, $0.30$~m and $0.50$~m (see Fig.~\ref{ExpSetup}b). These heights were chosen to coincide with the airflow zones within, at and above the calculated inner-layer depth (see below). Each anemometer array was sampled every $10$~s with data recorded by Campbell CR10X dataloggers. In analysis the velocity data were averaged over a $10$~minute period.

One anemometer array served as a reference site and was established permanently at the crest of the dune during the course of the experiment and at a distance of $\simeq 3.0$~m from the measurement transect (see Fig.~\ref{ExpSetup}a). In Fig.~\ref{ExpSetup}c, we show the velocity time series, corresponding to the upper ($0.5$~m) reference anemometer. We can see that during the $4$ hrs of the experiment the wind conditions at the reference site were relatively stable. Wind velocity at $0.5$~m height peaked at $12.1$~m s$^{-1}$, with an average of $7.7$~m s$^{-1}$ and an RMS value of $0.9$~m s$^{-1}$. The second (mobile) anemometer array was erected at each measurement site in turn for a period of $10$~minutes. Wind velocity data gained from the mobile array ($u_b$, $u_m$, $u_t$ from bottom to top respectively) were then normalised by data measured at the same height and time at the reference array ($U_b$, $U_m$, $U_t$) as a ratio (e.g. $u_b/U_b$).

Next to the reference anemometer array, a Sensit saltation impact sensor was established. The Sensit incorporates a piezo-electric transducer which records the impact of saltating sand grains as they bounce across the surface, thus providing data on the presence or absence of sand transport \cite{SZ97}. An example of the output from the Sensit is shown in Fig.~\ref{WindTransport}c. Data from the Sensit established that during the course of the experiment saltation transport occured for $\simeq 94 \%$ of the time. This suggests that the wind speeds we were measuring were above the critical entrainment threshold for erosion and therefore were significant for the evolving dynamics of the dune.

\section{Measuring the basal shear stress using time-averaged wind velocity: relevant time and length scales}
\label{timescales}

In both wind tunnel and atmospheric turbulent flows, the velocity $u$, averaged over the relevant time defined below, increases logarithmically with vertical distance from the surface $z$:
\begin{equation}
u=\frac{u_*}{\kappa} \ln \left(\frac{z}{z_0} \right),
\label{uzero}
\end{equation}
where $\kappa \simeq 0.4$ is the von K\'arm\'an constant; $u_*$ is the shear velocity, corresponding to a basal shear-stress $\rho_f u_*^2$, where $\rho_f$ is the air density; the aerodynamic roughness, $z_0$, due to the presence of a transport layer or of aeolian ripples, is typically of the order of a fraction of a millimeter. In accordance with a wind velocity of the order of $8$~m s$^{-1}$ at $z=0.5$~m (Fig.~\ref{ExpSetup}c), the shear velocity evident in this study was approximately $0.5$~m s$^{-1}$.

Importantly, the turbulent mixing time-scale at height $z$ is proportional to the inverse of the velocity gradient, and therefore scales as $z/u_*$. This sets the time-scale over which the measurement of $u$ should be averaged to effectively get the profile (\ref{uzero}) from the surface to a height $z$. For example, the logarithmic profile establishes over a thickness of $\sim 10$~cm above the surface in a fraction of second. In wind tunnels, the integral turbulent time-scale is of the order of $1$~s, and therefore the shear velocity extracted from the velocity profile for $z<10$~cm, averaged at that time-scale, is well defined. However, the integral turbulent time-scale of atmospheric flows is much longer, typically around $10$ to $15$~minutes. $u_*$ similarly extracted from Eq.~\ref{uzero} is then found to be time dependent. This shear velocity is recovered from the turbulent velocity fluctuations $(u',v',w')$ (at the Kolmogorov scale), measured at a centimetric distance to the bed, by averaging $\left< u'w' \right>$ over a time-window of the order of $1$~s.

\begin{figure}[t!]
\centerline{\includegraphics{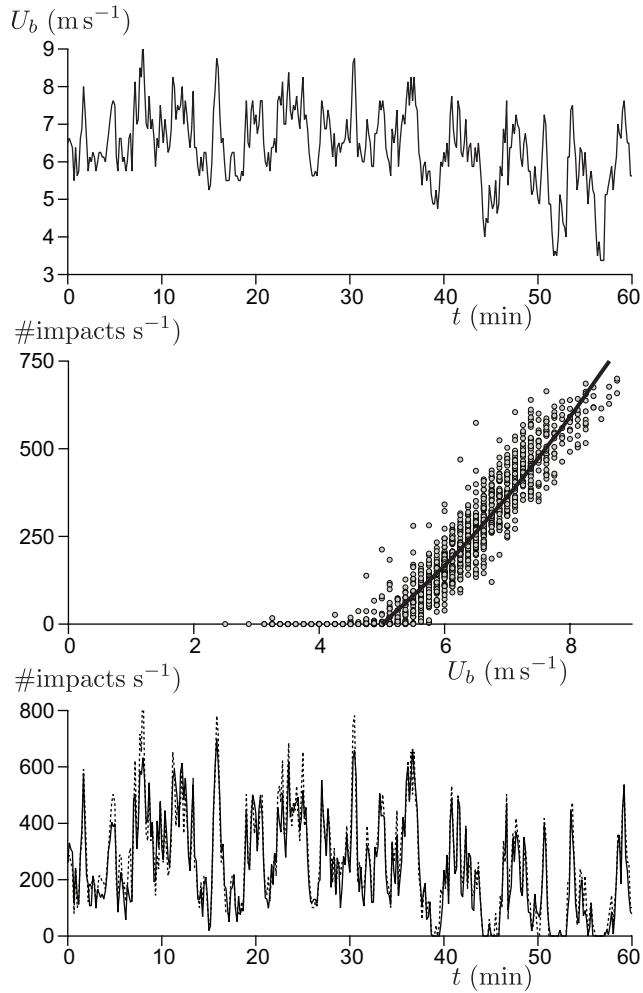}}
\caption{(a) Raw velocity data (bottom reference anemometer) sampled every $10$~s and measured over one hour. (b) Relationship between the grain impact counts per second, averaged over $10$~s, and the corresponding flow velocity $U_b$. Solid line: fit with a quadratic law $\propto (U_b^2 - U_{\rm th}^2)$. (c) Raw saltation data (solid line) compared to the above prediction using the corresponding wind velocity time series (dotted line).}
\label{WindTransport}
\end{figure}

Sediment transport (here saltation) has its own time scale: because of grain inertia, the grain flux responds to a change of flow velocity over a time scaling as $\frac{\rho_p}{\rho_f} \, \frac{d}{u_*}$, where $\rho_p$ is the density of the grains, which gives, here, an order of magnitude of $1$~s \cite{DCA11}. The coincidence of this time-scale with that of the wind tunnels makes the observed relationship between sediment flux and shear velocity well defined. With field data, a good correlation between transport and wind can also be found (Fig.~\ref{WindTransport}) when all quantities are measured at the relevant time and length scales: here both anemometer and grain impact sensor are located a few centimetres above the surface, and their signal is averaged over $10$~s.

Finally, over a dune or a gentle relief, the logarithmic velocity profile only holds in the so-called inner layer \cite{JH75}. An estimation of the thickness $\ell$ of this layer is related to the horizontal wavelength $\lambda$ of the relief by the scaling law \cite{TMB87}:
\begin{equation}
\frac{\ell}{\lambda} \frac{1}{\kappa^2} \ln^2\frac{\ell}{z_0} = \mathcal{O}(1).
\label{ell}
\end{equation}
We estimate the thickness $\ell$ for the dune under study ($\lambda \simeq 46$~m, see below) in the range $15$--$20$~cm.  As a consequence, the height at which the velocity is measured must be within this layer in order to extract the time-dependent $u_*(t)$ relevant to sediment transport. This is the case for our bottom anemometer, placed at $z=11$~cm.

\section{Results and analysis}
\label{results}

\begin{figure}[t]
\centerline{\includegraphics{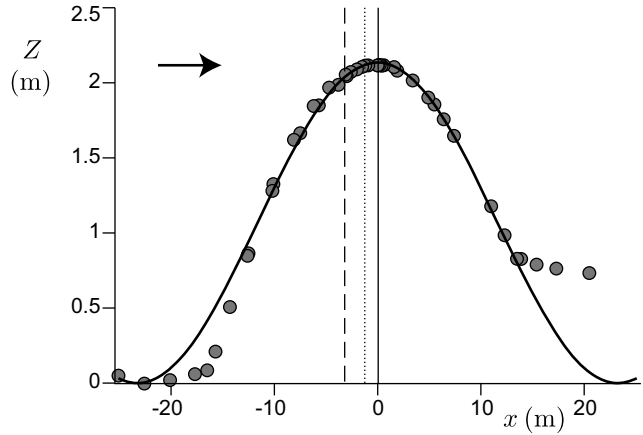}}
\caption{Dune elevation profile $Z(x)$. Wind is blowing from left to right (arrow). Symbols: measured data. Thick solid line: sinusoidal fit (Eq.~\ref{Zprofile}). Several vertical lines are displayed for reference in all figures. Solid line: position of dune crest ($x=0$). Dashed line: position of maximum velocity close to the surface (see Fig.~\ref{VelocityB}). Dotted line: position of vanishing erosion rate (see Fig.~\ref{Eros}).}
\label{ElevationProfile}
\end{figure}

The elevation profile of the dune on which wind velocity and erosion rates were measured is shown in Fig.~\ref{ElevationProfile}. Although the entire surface profile is not sinusoidal, the shape of the dune itself can be approximated by
\begin{equation}
Z(x) = Z_{\rm ref}+ \zeta \cos(kx),
\label{Zprofile}
\end{equation}
where $x=0$ is set at the position of the dune crest. The arbitrary reference level $Z_{\rm ref}$ is here chosen such that $Z=0$ at the upstream foot of the dune. Fitting this function to the data gives an effective wavenumber $k \simeq 0.14$~m$^{-1}$ (i.e. a wavelength $\lambda = 2\pi/k \simeq 46$~m) and an amplitude $\zeta \simeq 1.1$~m. The dune aspect ratio $2\zeta/\lambda \simeq 0.05$ means the topography is approximately at the upper limit for the hydrodynamical linear regime, which we investigate here, to be valid. A more refined analysis would involve the computation of the Fourier transform of the dune profile, in order to account for a whole range of wavenumbers $k$. This is, however, beyond the scope of this paper and would barely affect the results. For the sake of mathematical simplicity we therefore continue with a single mode analysis.

\begin{figure}[t]
\centerline{\includegraphics{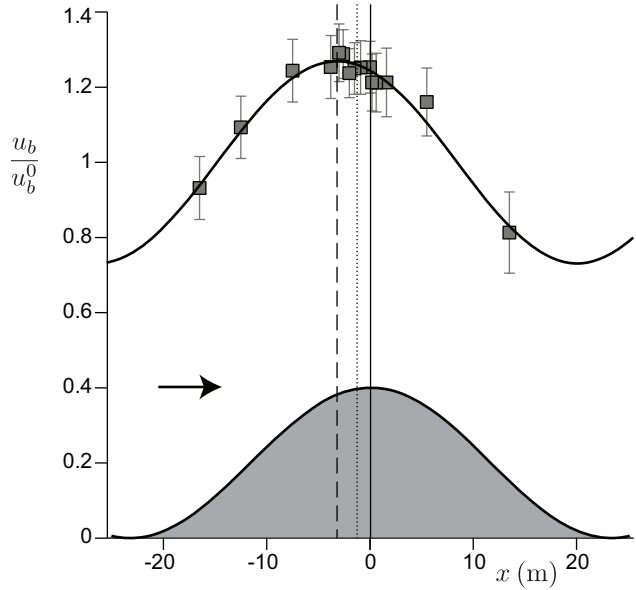}}
\caption{Longitudinal profile of the wind velocity $u_b$ at $11$~cm above the surface. Symbols: measured data. Thick solid line: sinusoidal fit (Eq.~\ref{uprofile}). Vertical reference lines: same legend as in Fig.~\ref{ElevationProfile}. In particular, the position at which the fitted line is maximum is marked with the dashed line. Error bars correspond to velocity RMS values. Bottom: sketch of the dune topography.}
\label{VelocityB}
\end{figure}

We display in Fig.~\ref{VelocityB} the longitudinal profile of the relative wind velocity measured with the bottom anemometer at $11$ centimeters above the surface, i.e. \emph{within the inner layer}. Consistently, with the linear analysis, this profile can also be fitted by a sinusoidal function of the same wave number $k$ as for the dune elevation, i.e. of the form:
\begin{equation}
u_b(x) = u_b^0+ \delta u_b \cos(kx+\varphi_b).
\label{uprofile}
\end{equation}
Importantly, $Z$ and $u_b$ are not in phase: the velocity reaches its maximum upstream of the crest. Here the phase difference is $\varphi_b \simeq 0.43$, i.e. around $26^\circ$, or $\varphi_b/k \simeq 3.2$~m upwind (vertical dashed lines in figures). For the fixed value of $k = 0.14$~m$^{-1}$ determined from the fit of the dune elevation (Fig.~\ref{ElevationProfile}), the accuracy on $\varphi_b$ is of the order of a few degrees. The two other fitting parameters are $u_b^0/U_b \simeq 0.8$ and $\delta u_b/u_b^0 \simeq 0.27$.

As discussed in the previous section, because the logarithmic law of the wall (Eq.~\ref{uzero}) locally holds in the inner layer at each position $x$, the velocity can be used as a proxy to calculate the basal shear stress with $\tau_b \propto \rho_f u_b^2$. At the linear order, we can write:
\begin{equation}
\tau_b(x) \propto \rho_f (u_b^0)^2 \left \{ 1 + k\zeta \left [ \mathcal{A}\cos(kx) - \mathcal{B}\sin(kx) \right ] \right \},
\label{taubprofile}
\end{equation}
where $\mathcal{A}$ and $\mathcal{B}$ are given by
\begin{equation}
\mathcal{A} = 2 \frac{\delta u_b}{u_b^0} \cos\varphi_b \frac{1}{k\zeta}
\quad \mbox{and} \quad
\mathcal{B} = 2 \frac{\delta u_b}{u_b^0} \sin\varphi_b \frac{1}{k\zeta} \, .
\label{AandBfromVel}
\end{equation}
Expression~(\ref{taubprofile}) is formally very general, while (\ref{AandBfromVel}) is specific to the approximation used here to relate $\tau_b$ to $u_b$. In fact, since sediment transport is controlled by the basal shear stress, these two parameters $\mathcal{A}$ and $\mathcal{B}$ are all that are needed as hydrodynamical inputs in bedform models, and are therefore of fundamental importance for the understanding of the mechanisms of dune initiation, instability and growth \cite{ECA05}.

With the different values of the parameters deduced from the fitting of the elevation and velocity profiles, we numerically get $\mathcal{A} \simeq 3.4\pm0.4$ and $\mathcal{B} \simeq 1.55\pm0.05$.  Error bars result from the account of uncertainty in the determination of these parameters, and of the selected wavenumber $k$ in particular. These values can be compared to linear theories that make predictions for $\tau_b$. They can be computed in the case of a semi-infinite turbulent flow over a wavy surface, with a first order turbulent closure on the mixing length $L$ of the rough type $L = \kappa(z+z_0)$ \cite{FCA10}. $\mathcal{A}$ and $\mathcal{B}$ are then found to be very weak (logarithmic) functions of $kz_0$. In our case, $kz_0$ is in the range $10^{-5}$--$10^{-4}$, which gives $\mathcal{A} \simeq 4$ and $\mathcal{B} \simeq 2$, i.e. close to our measured values. Besides, \cite{AXT94} have shown that the prediction of the basal shear stress is very robust with respect to the (arbitrary) choice of the turbulent closure. Finally, notice that non-linearities are expected to reduce this shift.

\begin{figure}[t]
\centerline{\includegraphics{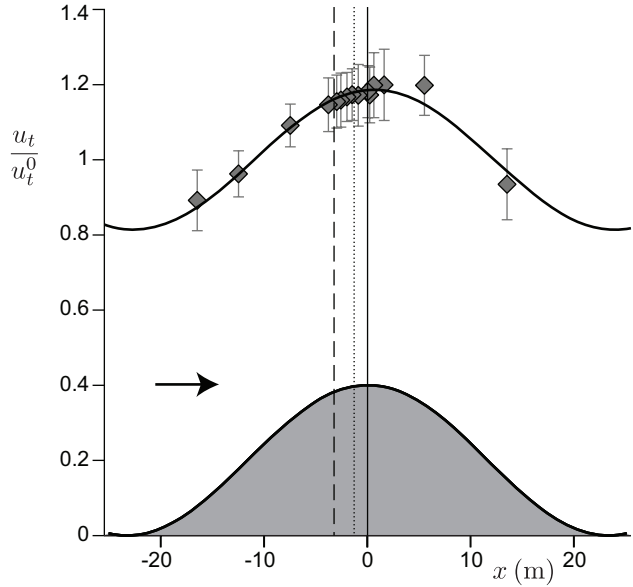}}
\caption{Longitudinal profile of the wind velocity $u_t$ at $50$~cm above the surface, i.e. above the inner layer. Symbols: measured data. Thick solid line: sinusoidal fit. Vertical reference lines: same legend as in Fig.~\ref{ElevationProfile}. Bottom: sketch of the dune topography.}
\label{VelocityT}
\end{figure}

We now focus on the effect of the vertical distance from the surface at which velocity is measured. In Fig.~\ref{VelocityT}, we display the profile of the wind velocity measured with the upper anemometer located at $50$ centimeters above the surface ($u_t$). In contrast with the profile evident with the bottom
anemometer ($u_b$; Fig.~\ref{VelocityB}), which is phased advanced with respect to topography, the top one has a small downwind phase shift ($\varphi_t \simeq -5^\circ$, or $0.6$~m downwind). The velocity profile from the middle anemometer $u_m$,  located at $30$ centimeters above the surface, is intermediate (not shown), with no detectable phase shift. The top anemometer is clearly located in the outer layer, where the dominant hydrodynamical mechanisms at work are different from those in the inner layer. In particular, rapid distortion and streamline curvature become important, and anisotropic second order turbulent closures are necessary to describe the flow in this region \cite{AXT94, FRBA90}. We interpret this slight phase delay in the outer layer to be related to the lag between production and dissipation of turbulent fluctuations \cite{WHCWWLC91, WLW96, vBAvD99, WN03}.

\begin{figure}[t]
\centerline{\includegraphics{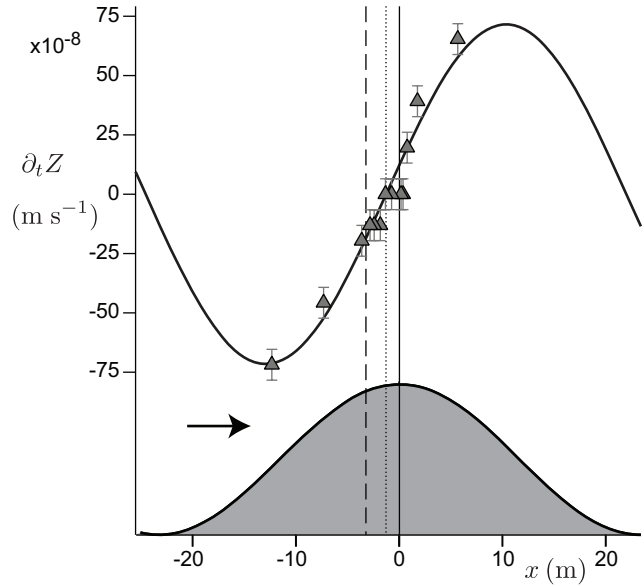}}
\caption{Longitudinal profile of the erosion rate $\partial_t Z$. Symbols: data measured with erosion pins. Thick solid line: sinusoidal fit. Vertical reference lines: same legend as in Fig.~\ref{ElevationProfile}. In particular, the position at which the fitted line vanishes is marked with the dotted line. Bottom: sketch of the dune topography.}
\label{Eros}
\end{figure}

Measured data for the erosion rate during the experiment are shown in Fig.~\ref{Eros}. In accordance with the linear analysis, we expect this profile to have longitudinal variations as  $\sin(kx+\varphi_q)$. Erosion rate and (volumetric) sand flux $q(x)$ are related through the matter conservation equation as $\partial_t Z + \partial_x q = 0$. The position at which $\partial_t Z$ vanishes then corresponds to the flux maximum. Here we find $\varphi_q \simeq 10^\circ$ (or $1.2$~m upwind, vertical dotted lines in figures), i.e. the maximum in sand flux is phase advanced to topography, but not as much as the near-surface velocity ($u_b$).

The reason why sediment flux is delayed with respect to the near-surface velocity is that sand transport does not adjust instantaneously to changes in shear stress: $q$ relaxes towards its equilibrium value $q_{\rm sat}$ over a spatial scale $L_{\rm sat}$, the so-called saturation length \cite{SKH01,ACD02,A04,ACP10}. These data then allow us to give an indirect estimate of $L_{\rm sat}$. Following our previous studies \cite{ACD02,ECA05,CA06,ACP10}, we assume for this purpose a first order (exponential) behavior for this relaxation $L_{\rm sat} \partial_x q = q_{\rm sat} - q$, with a transport law of the form $q_{\rm sat} = \chi (\tau_b - \tau_{\rm th})$, where $\tau_{\rm th}$ is the shear threshold for transport and $\chi$ a dimensionful constant. We then get:
\begin{equation}
kL_{\rm sat} = \tan(\varphi_b-\varphi_q) = \frac{\mathcal{B}-\mathcal{A}\tan\varphi_q}{\mathcal{A}+\mathcal{B}\tan\varphi_q} \, .
\label{Lsatfromphiq}
\end{equation}

With the above values of $\mathcal{A}$, $\mathcal{B}$ and $k$, we compute $L_{\rm sat} \simeq 2$~m. Taking into account the fact that the shear threshold depends on the bed slope as $\tau_{\rm th} = \tau_{\rm th}^0 (1+\partial_x Z/\mu)$, where $\mu \simeq 0.6$ is the avalanche slope, Eq.~\ref{Lsatfromphiq} is modified with $\mathcal{B}$ replaced by $\mathcal{B}_\mu=\mathcal{B}-\frac{1}{\mu} \frac{\tau_{\rm th}^0}{\tau_b^0}$. Here we take $\tau_{\rm th}^0/\tau_b^0 \simeq (u_{\rm th}^0/u_b^0)^2$, a ratio estimated around $0.4$ during our experiment, i.e. $\mathcal{B}_\mu \simeq 0.9$. This leads to  $L_{\rm sat} \simeq 0.7$~m. Previous direct measurements of the saturation length performed by \cite{ACP10} gave $L_{\rm sat} \simeq 0.55 \pm 0.1$~m for grains of diameter $d \simeq 120$~$\mu$m (wind tunnel experiments), and $L_{\rm sat} \simeq 1 \pm 0.2$~m for grains of diameter $d \simeq 185$~$\mu$m (field experiments). Here, as $d \simeq 310$~$\mu$m and since the saturation length is expected to be proportional to the grain size ($L_{\rm sat} \propto d \rho_p/\rho_f$), the range into which this indirect estimate falls is fair, given the relatively low spatial resolution ($0.5$~m at most) with which the erosion rate profile was measured.

\section{Conclusions}
\label{conclu}

Whilst the results shown here are based on measurements undertaken on a single dune and more data are required on differing dune sizes and with different wind velocities, our data do provide the first field evidence of the upwind phase shift of velocity close to the surface with respect to topography. Importantly, this shift is only observed when the velocity is measured in the inner layer: it vanishes and even changes sign (velocity maximum downwind of the dune crest) when the measurement is undertaken at greater heights above the surface. The value of this upwind phase-shift is, furthermore, in quantitative agreement with the prediction of the hydrodynamical linear analysis with first order closures for turbulent flows. In conclusion, the two ingredients of the dune instability mechanism that are required to explain sand dune initiation and growth, namely (i) the upwind phase-shift of the basal shear stress with respect to the topography and (ii) the spatial lag $L_{\rm sat}$ of sand transport with respect to this stress \cite{ACP10}, are now both experimentally validated for the first time. This now provides the opportunity for far more confident modelling of sand dune dynamics.

Previous field tests of the predictions of \cite{JH75} have been performed on large hills \cite{TMB87}, making the entire longitudinal profile of the velocity close to the surface difficult to record. As a result, only the overall speed-up and the general structure of the wind flow has been assessed. In the other extreme, wind tunnel studies on decimeter-scale bumps do not allow access to the inner layer, which is too small in this case. Here, we have illustrated that dunes provide good intermediate topographies over which turbulent predictions can be tested in detail.

\vspace*{0.3cm}
\noindent
\rule[0.1cm]{3cm}{1pt}

BA and PC are grateful to F. Charru for stimulating discussions. The help of H. Elbelrhiti, L. Kabiri and L. Olver has been very welcome for the field work. We thank ANR Zephyr grant $\#$ERCS07\underline{\ }18 for funding.


\end{document}